\documentclass[12pt]{article}
\usepackage{amssymb}
\usepackage{amsmath}
\usepackage{amsfonts}
\newcommand{\sign}{\mathop{\rm sign} \nolimits}
\renewcommand{\det}{\mathop{\rm det} \nolimits}

\makeatletter
\renewcommand{\section}{\@startsection{section}{1}{0pt}%
{3.5ex plus 1ex minus .2ex}{2.3ex plus .2ex}%
{\large\bf}}
\makeatother
\begin{document}

\begin{center} {\bf ON DYNAMICS OF STRINGS AND BRANES}
\end{center}

\begin{center} {A.V. Golovnev, L.V. Prokhorov\\
{\it \small V.A. Fock Institute of Physics \\ Sankt-Petersburg State University}}\end{center}

\begin {center} {\bf Abstract} \end{center}

We study Nambu-Goto strings and branes.  It is shown that they can be considered
as continuous limits
of ordered discrete sets of relativistic particles for which the tangential
velocities are excluded from the action. The linear in unphysical momenta 
constraints are found. It allows to derive the
evolution operators for the objects under consideration
from the "first principles".

\section {Introduction.}
We consider Nambu-Goto strings and branes. We start with the simplest case of one
relativistic particle which formally can be considered as a 0-brane. The relativistic
invariant form of its action is raparametrization invariant and has a constraint,
commonly taken in a quadratic form with the loss of information on the
momentum sign. It led to some problems [1,2] in quantum description of 
free relativistic particles.
In section 2 we explain how these problems can be overcome [3,4] using a linear "constraint"
containing the sign of the velocity $\frac{\partial x^0}{\partial\sigma_0}$ \
($\frac{\partial t}{\partial\tau}$ in usual notations) and fixing the physical sector of
quantum theory by the condition $\frac{\partial x^0}{\partial\sigma_0}>0$.
After that the operator of evolution (and propagator [3])
can be derived from the "first principles"
of quantum mechanics.

In sections 3 and 4 we prove that p-branes may be considered as continuous 
limits of discrete sets of relativistic
particles and that the $p$-brane action
\begin{equation}
\label{Nambu-Goto}
S=
-\gamma\int d^{\ p+1}\sigma\sqrt{(-1)^p\ g},\ \ g=
\det\frac{\partial x^{\mu}}{\partial\sigma_a}%
\frac{\partial x_{\mu}}{\partial\sigma_b},
\end{equation}
is a continuous limit of sum of properly modified
relativistic particle actions.  
Here $a,b=0,1,\ldots,p$ and $\mu=0,1,\ldots,n$ where $p+1$ and 
$n+1$ are the brane worldsheet and the bulk space dimentions respectively;
$g$ is the induced metric determinant on the worldsheet.
Strings correspond to the $p=1$ case in (\ref{Nambu-Goto}).
The modification is such that particle motions along
the brane hypersurface become unphysical. It yields $p$ constraints; the remaining constraint $(H=0)$
is a consequence of arbitrariness of "time" $\sigma_0$.

In sections 5 and 6 we consider dynamics of
strings ($p=1$) and branes ($p>1$).
The theories are reparametrization invariant for all $p$, $n>p$; each of them has
$p+1$ constraints. Usualy one of these constraints physicists take
in a quadratic form. It causes the same problems as in the case of a single particle. We 
found that the operator of evolution can be constructed in the
same way as in section 2; new difficulties are rather technical than conceptual ones.
 The solution is also similar
to the $p=0$ case (a particle).

\section {Relativistic particle.}
Motion of free relativistic particle is defined by the well-known action
$$S=-m\int\sqrt{1-{\overrightarrow v}^2}dt,$$ 
where $\overrightarrow v=\frac{d\overrightarrow x(t)}{dt}$. The canonical momentum is
$\overrightarrow{\mathfrak p}\equiv\frac{\partial L}{\partial\overrightarrow v}=
\frac{m\overrightarrow v}{\sqrt{1-{\overrightarrow v}^2}}\equiv E_{\mathfrak p}\overrightarrow v$
with $L$ being the Lagrangian, and the Hamiltonian is $H=E_{\mathfrak p}=\sqrt{m^2+
{\overrightarrow{\mathfrak p}}^2}$. We can write down the action in the explicitly  relativistic
invariant form by parametrization of the world line: $x^{\mu}=x^{\mu}(\sigma_0)$ 
(usually one uses $\tau$ instead of $\sigma_0$) with $x^{\mu}(\sigma_0)=
(t(\sigma_0),\overrightarrow x (\sigma_0)), \ \mu=0,\ldots, n$.
We denote $\dot x^{\mu}\equiv\frac{dx^{\mu}}{d\sigma_0}$, so that
$\overrightarrow v=\dot{\overrightarrow x}\ \frac{d\sigma_0}{dt}$ and 
$$S=-m\int\sqrt{{\dot x}^{\mu}{\dot x}_{\mu}}%
d\sigma_0.$$ It is the $p=0$ case of (\ref{Nambu-Goto}) in which we put
$m$ (particle's mass) instead of $\gamma$.
The vector of canonical momentum is
\begin{equation}
\label{momentum}
{\mathfrak p}_{\mu}\equiv\frac{\partial L}%
{\partial {\dot x}^{\mu}}=-m\frac{{\dot x}_{\mu}}{\sqrt{{\dot x}^2}}.
\end{equation}

The obtained theory is invariant under reparametrization group
$\sigma_0\rightarrow \tilde\sigma_0=%
f(\sigma_0)$),
hence its Hamiltonian is zero and by squaring the equation (\ref{momentum})
one gets a constraint:
\begin{equation}
\label{partconst}
{\mathfrak p}^2-m^2=0.
\end{equation}
 Of course, the information about
the sign of ${\mathfrak p_0}$ is lost. On the other hand one can prove that
any constraint has to be linear in unphysical momenta [3,4]. It was a serious obstruction
to the {\it ab initio} derivation of relativistic particle propagator
(authors of [1,2] used some additional assumptions).
The problem was overcome only in [3]. One can obtain the solution in the
following way.

From the constraint (\ref{partconst}) one finds ${\mathfrak p}_0=\pm%
\sqrt{m^2+{\overrightarrow {\mathfrak p}}^2}$, and it is obvious from (\ref{momentum}) 
that $\sign {\mathfrak p}_0=-\sign \dot x_0$.
 Combining these facts we have 
\begin{equation}
\label{constraint}
{\mathfrak p}_0+E_{\mathfrak p}\sign({\dot x}^0)=0
\end{equation}
with $E_{\mathfrak p}({\overrightarrow {\mathfrak p}})=\sqrt{m^2+{\overrightarrow {\mathfrak p}}^2}$.
The Hamiltonian is zero and the
total Hamiltonian [5] is
 $$H_T=v\left ({\mathfrak p}_0+E_{\mathfrak p}\sign({\dot x}^0)\right ).$$ 
Here $v$ is the Lagrange multiplier. Strictly speaking Eq. (\ref{constraint}) is not
a constraint because it contains velocity
(and $H_T$ is not a Hamiltonian due to the same reason). But it depends only upon the
sign of $\dot x$, and this fact allows us to formulate the quantum theory.

We fix the $\sigma_0$ "time arrow" by condition
\begin{equation}
\label{condition}
\frac{\partial x^0}{\partial\sigma_0}>0
\end{equation}
which forces us to admit that
$v>0$ (because $\Delta x^0=v\Delta\sigma_0$, see later). 
We use the evolution operator $$U_{\omega}(x, \tilde x)=%
\left\langle x|\exp (-i\omega \hat H_T)|\tilde x\right\rangle,$$  
$$\psi(x, \sigma_0+\omega)=\int d^4\tilde x\, U_{\omega}(x, \tilde x)%
\psi(\tilde x, \sigma_0)$$ and the following relation
for infinitesimal $\omega$:
$$\left\langle x|\exp (-i\omega \hat H_T)|\tilde x\right\rangle=
\int d^4{\mathfrak p} \left\langle x|\exp (-i\omega H_T(x,{\mathfrak p}))|{\mathfrak p}\right\rangle%
\left\langle {\mathfrak p}|\tilde x\right\rangle .$$
Integrating over ${\mathfrak p}_0$ and
 (with use of $\delta$-function $\delta (x^0-\tilde x^0-\omega v)$) over $\tilde x_0$
 we get for the wave function [4]:
\begin{equation}
\label{evolution}
\psi(x^0, \overrightarrow x)=\int\frac{d^3{\mathfrak p}d^3\tilde x}{(2\pi)^3}%
\exp\left (i[{\mathfrak p}_i\Delta x^i-\Delta x^0E_{\mathfrak p}]\right )\psi ({\tilde x}^0, %
\overrightarrow{\tilde x}),
\end{equation} 
where $\Delta x^0=v\omega$, and we omit the argument $\sigma_0$ because all the
information on $\sigma_0$ is accumulated in $x^0$. Using 
(\ref{evolution}) one can get the right Feynman propagator for a relativistic particle
[4].

\section {String as an ordered system of relativistic particles.}
Now we show that any string and brane can be described as a system of particles. More precisely, 
the action (\ref{Nambu-Goto}) may be regarded as a continuous limit of
sum of free relativistic particle actions provided that we take
$\overrightarrow v_{\perp}$ instead of $\overrightarrow v$, where
$\overrightarrow v_{\perp}$ is the part of velocity orthogonal to the constant time hypersurface
of the brane worldsheet. 
We deal here with a kind of indirectly introduced particle
interaction.

In this section we consider a string (1-brane) and reproduce the proof of our statement by
an explicit calculation [6].  We consider $N+1$ particles
with the position vectors $\overrightarrow x_k(x^0)$,\ $k=0,1,\ldots,N$ and the action
\begin{equation}
\label{discr}
S=-m\sum\limits_{k=0}^N\int dx^0\sqrt{1-\overrightarrow v_{k \perp}^2(x^0)},
\end{equation}
in which $\overrightarrow v_k=\frac{d\overrightarrow x_k}{d x^0}$ and 
$$\overrightarrow v_{k \perp}=\overrightarrow v_k-
\frac{\left (\overrightarrow v_k \cdot \Delta\overrightarrow x_k\right )}
{(\Delta\overrightarrow x_k)^2}\Delta\overrightarrow x_k,\ \ 
\Delta\overrightarrow x_k\equiv\overrightarrow x_{k+1}-\overrightarrow x_k.$$
In the continuous limit we define $\frac{kA}{N}\rightarrow\sigma_1,\  
\frac{\Delta\overrightarrow x_k}{A/N}\rightarrow\overrightarrow k(\sigma_1),\ 
\frac{m}{|\Delta\overrightarrow x_k|}\rightarrow\gamma$.
Here $\sigma_1\in[0,A]$; usually one takes $A=\pi$, but for our purposes it may be natural 
to consider $A$ as the string length and $|\Delta\overrightarrow x_k|=\frac{A}{N},\ 
|\overrightarrow k|=1$.
In any case we have
\begin{equation*}
S=-\gamma\sum_{k=0}^{N}\int dx^0\Delta l\sqrt{1-{\overrightarrow v}_{k\perp}^2}
\to
-\gamma\int dx^0 |\overrightarrow k|d\sigma_1\sqrt{1-{\overrightarrow v}_{\perp}^2}
\end{equation*}
with $\overrightarrow k=\frac{\partial {\overrightarrow x}(x^0,\sigma_1)}{\partial\sigma_1}$, \ 
$\overrightarrow v=\frac{\partial {\overrightarrow x}(x^0,\sigma_1)}{\partial x^0}$ and
$\overrightarrow v_{\perp}=\overrightarrow v-
\frac{(\overrightarrow v\overrightarrow k)}{k^2}\overrightarrow k$.
The string length is $L=\int|\overrightarrow k|d\sigma_1$.
After that we parametrize the worldsheet $x^0=x^0(\sigma_0,\sigma_1),\ 
\overrightarrow x=\overrightarrow x(\sigma_0,\sigma_1)$ introducing a new
parameter $\sigma_0$
(again the standard notations are $x^0\rightsquigarrow t$ and $\sigma_0\rightsquigarrow\tau$). 
We have $\dot x\equiv\frac{\partial x(\sigma_0,\sigma_1)}{\partial\sigma_0}
=(1,\overrightarrow v)\dot x_0,$ \
${x}^{\prime}\equiv\frac{\partial x(\sigma_0,\sigma_1)}{\partial\sigma_1}
=({x_0}^{\prime},\overrightarrow k +\overrightarrow v {x_0}^{\prime})$ 
and
\begin{equation*}
S=
-\gamma\int d^2\sigma\dot x^0 |\overrightarrow k|\sqrt{1-{\overrightarrow v}_{\perp}^2}.
\end{equation*}
The last expression is equal to the Nambu-Goto action:
\begin{equation*}
S=
-\gamma\int d^2 \sigma \sqrt{(\dot x x^{\prime})^2-{\dot x}^2{x^{\prime}}^2}.
\end{equation*}
Of course, one could start with it and get $S=-\gamma\int dx^0dl\sqrt{1-{\overrightarrow v_{\perp}}^2}$ which is a continuous limit
of (\ref{discr}).

We can also propose another discrete analogue of the Nambu-Goto string. Let's consider
the following action: $$S=-m\sum\limits_{k=0}^N\int d\sigma_0\sqrt{\dot x^{\mu}_{k\perp}
\dot {x_{\mu}}_{k\perp}},$$ where $\dot x^{\mu}_{k\perp}$ is the part of $\dot x^{\mu}_k$
perpendicular to $x^{\mu}_{k+1}-x^{\mu}_{k}$. The continuous limit is $N\to\infty,\ 
\frac{kA}{N}\to\sigma_1,\ \frac{m}{|\Delta s_k|}\to\gamma$ with the invariant interval
$(\Delta s_k)^2=(x^{\mu}_{k+1}-x^{\mu}_{k})({x_{\mu}}_{k+1}-{x_{\mu}}_{k})$. We have
$\dot x^{\mu}_{\perp}=\dot x^{\mu}-\frac{\dot x^{\nu}x^{\prime}_{\nu}}%
{x^{\prime\rho}x^{\prime}_{\rho}}x^{\prime\mu}$,
$\left (\frac{ds}{d\sigma_1}\right )^2=x^{\prime 2}$,
 and
$$S=-\gamma\int d\tau |ds|\sqrt{{\dot x}^2-
\frac{(\dot x x^{\prime})^2}{{x^{\prime}}^2}}=-\gamma\int d\sigma_0 d\sigma_1
\sqrt{(\dot x x^{\prime})^2-{\dot x}^2{x^{\prime}}^2}.$$ In contrast to the previous paragraph
the presented discrete theory has the relativistic invariant form from
the very begining but even the sense of $\dot x_{\perp}$ depends upon
the parametrization of the worldsheet. In the gauge
$\sigma_0=x^0$ these two approaches coincide.

\section {Brane as an ordered system of relativistic particles.}
Now we shall prove our statement for $p>1$. We consider $(N+1)^p$
particles arranged into some $p$-dimensional lattice with 
the position vectors $\overrightarrow x_{i_1 i_2\ldots i_p}$ and
the action
$$S=-m\int dx^0\sum\limits_{i_1=0}^{N}\cdots\sum\limits_{i_p=0}^{N}\sqrt{1-
\overrightarrow v^2_{i_1,\ldots i_p\perp}}$$ where $\overrightarrow v_{i_1,\ldots i_p\perp}$
is the component of $\overrightarrow v_{i_1\ldots i_p}$ perpendicular to
$\overrightarrow x_{i_1\ldots i_k+1\ldots i_p}-\overrightarrow x_{i_1\ldots i_k\ldots i_p}$
for all $k$. In continuous limit we demand $\frac{Ai_k}{N}\to\sigma_k$ and
$\frac{m}{\Delta V}\to\gamma$ with $\Delta V$ being volume of a cell of the lattice.
The action takes the form $S=-\gamma\int dx_0 dV\sqrt{1-{\overrightarrow
 v}_{\perp}^2}$, and we need to prove that $S$ is equal to (\ref{Nambu-Goto}).

For the sake of simplicity first we consider some special coordinate system
on the brane. Let the brane be parametrized by $\sigma_0, \sigma_1,\ldots,\sigma_p$, \quad
$\sigma_i\in[0,A]$ for 
$i=1,\ldots,p$. We choose a coordinate system in which $\sigma_0=x^0$,
so that $\frac{\partial x^0}{\partial\sigma_i}=0$,
 and
$\frac{\partial x^{\mu}}{\partial\sigma_i}\frac{\partial x_{\mu}}%
{\partial\sigma_j}=0, \ i\neq j$. It is always possible,
locally at least.
Let's denote
$\overrightarrow k_i\equiv%
\frac{\partial\overrightarrow x(x^0,\sigma_i)}{\partial\sigma_i}$ and $\overrightarrow v\equiv
\frac{\partial\overrightarrow x(x^0,\sigma_i)}{\partial x^0}$ on the worldsheet.
In our coordinate system
$\overrightarrow k_i$ is orthogonal to $\overrightarrow k_j$,
$\ i\neq j$ and $$\overrightarrow v_{\perp}=\overrightarrow v-\sum_{i=1}^{p}\frac{(\overrightarrow v\overrightarrow k_i %
)}{{k_i}^2}\overrightarrow k_i.$$
Of course, one can find $\overrightarrow v_{\perp}$ without the orthogonality condition
with the use of standard orthogonalization procedure, but in section 6
more simple proof is presented.

We have $\frac{\partial x^{\mu}}{\partial\sigma_0}=(1,\overrightarrow v)$ and $\frac{\partial x^{\mu}}{\partial\sigma_i}=(0,\overrightarrow k_i)$,
so the determinant in (\ref{Nambu-Goto}) is 
\begin{multline*}
\det \frac{\partial x^{\mu}}{\partial\sigma_a}\frac{\partial x_{\mu}}{\partial\sigma_b}=\left |
\begin{array}{ccccc}
1-{\overrightarrow v}^2 & -\overrightarrow v \overrightarrow k_1 &
-\overrightarrow v \overrightarrow k_2 & \ldots & -\overrightarrow v \overrightarrow k_p \\
-\overrightarrow v \overrightarrow k_1 & -k^2_1 & 0 & \ldots & 0 \\
-\overrightarrow v \overrightarrow k_2 & 0 & -k^2_2 & \ldots & 0 \\
\vdots & \vdots & \vdots & \ddots & \vdots \\
-\overrightarrow v \overrightarrow k_p & 0 & 0 & \ldots & -k^2_p
\end{array}
\right | =\\
=\left (\prod\limits_{i=1}^p (-{k_i}^2)\right )%
\left (1-{\overrightarrow v}^2+\sum\limits_{i=1}^p\frac{\left(\overrightarrow v %
\overrightarrow k_i\right )^2}{{k_i}^2}\right )
\end{multline*}
 where
$a,b=0,1,\ldots,p$. It is not also difficult to see that the volume element at the 
constant time hypersurface on the brane worldsheet is $dV=\prod\limits_{i=1}^p |k_i|d\sigma_i$. We conclude that the action
\begin{equation}
\label{setof}
S=-\gamma\int d\sigma_0 d^n\sigma_i\sqrt{\left|\det \frac{\partial x^{\mu}}{\partial\sigma_a}\frac{\partial x_{\mu}}{\partial\sigma_b}\right|}=-\gamma\int dx_0 dV\sqrt{1-{\overrightarrow
 v}_{\perp}^2}.
\end{equation}
It proves our statement.

\section {Dynamics of strings.}
As it was mentioned in the Introduction, the evolution operator $U_{\omega}$ can be constructed for strings and branes
in the same way as in section 2 [7].
For $p=1$ action (\ref{Nambu-Goto}) is the action of free bosonic string
with the Lagrange density [6,8-11]
\begin{equation}
\label{straction}
{\mathcal L}=-\gamma\sqrt{-g}=-\gamma\sqrt{(\dot x x^{\prime})^2-{\dot x}^2{x^{\prime}}^2}.
\end{equation}
We assume that $\dot x$ is timelike and $x^{\prime}$ is
spacelike, so that $\sigma_0$ can be regarded as a time
parameter. In this case the momentum 
\begin{equation}
\label{strmom}
{\mathfrak p}_{\mu}\equiv\frac{\partial\mathcal L}{\partial\dot x^{\mu}}=%
-\gamma\frac{(\dot x x^{\prime})x^{\prime}_{\mu}-%
{x^{\prime}}^2\dot x_{\mu}}{\sqrt{(\dot x x^{\prime})^2-%
{\dot x}^2{x^{\prime}}^2}}
\end{equation}
will also be timelike. One easily gets two constraints
\begin{eqnarray}
\label{constone}
{\mathfrak p}_{\mu}%
{x^{\prime}}^{\mu}=0,\\
\label{consttwo}
{\mathfrak p}^2+{\gamma}^2{x^{\prime}}^2=0.
\end{eqnarray}
The second one is obtained by squaring the Eq. (\ref{strmom}) and hence some information is lost.
As in the case of a pointlike particle Eq. (\ref{consttwo}) yields 
${\mathfrak p}_0=\pm E_{\mathfrak p}$ with $E_{\mathfrak p}=\sqrt%
{{\overrightarrow {\mathfrak p}}^2-{\gamma}^2{x^{\prime}}^2}$.  The sign of ${\mathfrak p}_0$
follows from the definition of the momentum. We have
 $${\mathfrak p}_0+E_{\mathfrak p}(x, \overrightarrow {\mathfrak p})\sign (y_0)=0,$$ where
$y_{\mu}=(\dot x x^{\prime})x^{\prime}_{\mu}-{x^{\prime}}^2\dot x_{\mu}$.
The vector $y$ is obviously timelike (indeed, 
$y^2={x^{\prime}}^2g>0$ and 
$y_{\mu}\dot x^{\mu}=-g>0$, so that $\sign \dot x^0=\sign y^0$).
We get the "constraint" \ analogous to (\ref{constraint}):
\begin{equation}
\label{strconst}
{\mathfrak p}_0+E_{\mathfrak p}(x, \overrightarrow {\mathfrak p})\sign (\dot x^0)=0.
\end{equation}

The \ "Hamiltonian" of the theory $H={\mathfrak p_0}+E_{\mathfrak p}\sign{\dot x}^0$ is zero, and demanding again $\sign \dot x^0>0$ we get the total Hamiltonian
$H_T=u({\mathfrak p}_0+E_{\mathfrak p})+v{\mathfrak p}_{\mu}{x^{\prime}}^{\mu}$. 
We have two unphysical momenta and
need to exclude them from $E_{\mathfrak p}$. 
Let's exclude ${\mathfrak p}_0$ and ${\mathfrak p}_1$ (we denote the remaining
components by the lower index "$_{>}$": \ ${\mathfrak p}^{\mu}=
({\mathfrak p}_0,{\mathfrak p}_1,{\mathfrak p}_{>})$).
One can find ${\mathfrak p}_1$ from (\ref{constone}) if $x^{\prime}_{1}\neq 0$. 
In [7] we restricted ourselves to the case
$x^{\prime}%
_0=0$ and had
$$E_{\mathfrak p}(x,{\mathfrak p}_{>})=
\sqrt{\left (\frac{{\mathfrak p}_{>}x^{\prime}_{>}}{x^{\prime}_{1}}\right )^2%
+{\mathfrak p}^2_{>}-\gamma^2{x^{\prime}}^2}.$$ In general case one can substitute 
${\mathfrak p}_1$
from (\ref{constone}) to (\ref{consttwo}) and get a quadratic equation for
${\mathfrak p}_0$:
$${{\mathfrak p}_0}^2\left (1-{\left (\frac{{x^{\prime}}^0}{{x^{\prime}}^1}\right )}^2
\right )+2{\mathfrak p}_0\frac{({\mathfrak p}_{>}x^{\prime}_{>}){x^{\prime}}^0}{{({x^{\prime}}^1)}^2}-
{\mathfrak p}_{>}^2\left (1+\left (\frac{x^{\prime}_{>}}{{x^{\prime}}^1}\right )^2\right )+
\gamma^2{x^{\prime}}^2=0.$$ If $|{x^{\prime}}^1|>|{x^{\prime}}^0|$, it has two 
real roots of opposite signs, and we can choose a proper one following (\ref{strconst}).
Otherwise we have to try to exclude another component of ${\mathfrak p}_{\mu}$ from
$E_{\mathfrak p}$. It's always possible because $x^{\prime}$ is spacelike and 
$|\overrightarrow {x^{\prime}}|>|{x^{\prime}}^0|$. In section 3 we have seen that the unphysical 
degree of freedom is related to the motion of particles along the string.

Now we  write down the
evolution equation (see section 2)
\begin{multline*}
\psi(x)=\int{\mathit D}^{n+1}\tilde x(\sigma_1)\left\langle x(\sigma_1)|\exp (-i\omega%
 \hat H_T)|\tilde x(\sigma_1)\right\rangle\psi(\tilde x(\sigma_1))=\\
=\int{\mathit D}^{n+1}{\mathfrak p}{\mathit D}^{n+1}\tilde x\exp(i[{\mathfrak p}_{\mu}%
\Delta x^{\mu}-\omega u({\mathfrak p}_0+E_{\mathfrak p}(\tilde x, {\mathfrak p}_{\perp}))-%
\omega v{\mathfrak p}_{\mu}{x^{\prime}}^{\mu}])\psi%
 (\tilde x)=\\
=\int{\mathit D}^{n-1}{\mathfrak p}_{>}{\mathit D}^{n+1}\tilde x\exp (i[-{\mathfrak p}_{>}\Delta x_{>}-%
\omega uE_{\mathfrak p}(\tilde x, {\mathfrak p}_{>})+\omega v{\mathfrak p}_{>}x^{\prime}_{>}])\times\\
\times\delta(\Delta x^0-\omega u-\omega v{x^{\prime}}^0%
)\delta(-\Delta x_1+\omega vx^{\prime}_1)\psi(\tilde x).
\end{multline*}
Here $D\tilde x$ and $D{\mathfrak p}$ denote differentials in the functional spaces
and all the integrals are functional ones (path integrals).

$\delta$-Functions determine the Lagrange multipliers $\omega v=\frac{\Delta x_1}%
{x^{\prime}_{1}}$ and $\omega u=\Delta x^0-\frac{{x^{\prime}}^0\Delta x_1}%
{x^{\prime}%
_1}$ yielding the final result
\begin{multline*}
\psi(x)=\int{\mathit D}^{n-1}{\mathfrak p}_{>}{\mathit D}^{n-1}{\tilde x}_{>}%
\exp(i[-{\mathfrak p}_{>}\Delta x_{>}-\\
-\left (\Delta x^0-\frac{{x^{\prime}}^0\Delta x_1}%
{x^{\prime}_1}\right )E_p(\tilde x, {\mathfrak p}_{>})+\frac{\Delta x_1}{x^{\prime}_1}%
{\mathfrak p}_{>}x^{\prime}_{>}])\psi(\tilde x).
\end{multline*}
If ${x^{\prime}}^0=0$, Eq. (\ref{condition}) implies $u>0$.

\section {Dynamics of branes.}
 We turn to the general case of action (\ref{Nambu-Goto}). We denote
the $\sigma_0, \sigma_1,\ldots,\sigma_p$ derivatives of $x$ by
$x_{,0}, x_{,1},\ldots,
x_{,p}$. Again we assume that the vector $x^{\mu}_{,0}$ is timelike, and vectors $x^{\mu}_{,i}$ 
are spacelike (here and hereafter in this chapter $i, k, l=1,\ldots ,p$ while $a, b=0,\ldots ,p$). 
Now in the action (\ref{Nambu-Goto}) 
\begin{equation*}
g(\sigma)=\frac{1}{(p+1)!}{\epsilon}_{a_0\ldots a_p%
}{\epsilon}^{b_0\ldots b_p} x_{\alpha_0,b_0}x^{\alpha_0,a_0}
\cdots x_{\alpha_p,b_p}x^{\alpha_p,a_p}
\end{equation*}
with $\epsilon$ being the unit antisymmetric Levi-Civita symbol, and the canonical momentum is
\begin{equation*}
{\mathfrak p}_{\mu}
=\frac{-\gamma (-1)^{p}}{p!\sqrt{(-1)^p g}}{\epsilon}_
{0a_1\ldots a_p}{\epsilon}^{b_0\ldots b_p}x_{\mu,b_0}\, x_{,b_1}\cdot x^{,a_1}
\ldots x_{,b_p}\cdot x^{,a_p}.
\end{equation*}
Evidently ${\mathfrak p}_{\mu}x^{\mu}_{,i}=0$ 
 due to antisymmetry of $\epsilon$, and using the equality
$$\epsilon^{a_0\ldots a_p}g(\sigma)=\epsilon^{b_0\ldots b_p}
x_{,b_0}\cdot x^{,a_0}
\ldots x_{,b_p}\cdot x^{,a_p}$$ one obtains
${\mathfrak p}^2=(-1)^{p}{\gamma}^2 \zeta (x),\ \ \zeta (x)=\det
x^{\mu}_{,i}x_{\mu ,k}$.
So, with the loss of information about the sign of ${\mathfrak p}_0$, the constraints are
\begin{eqnarray}
\label{brone}
{\mathfrak p}_{\mu}x^{\mu}_{,i}=0,\ \ \ i=1,2,\ldots,p;\\
\label{brtwo}
{\mathfrak p}^2-(-1)^p\gamma^2\zeta (x)=0.
\end{eqnarray}
From (\ref{brtwo})
 we have ${\mathfrak p}_0=\pm E_{\mathfrak p}$ with 
$E_{\mathfrak p}=\sqrt{{\overrightarrow {\mathfrak p}}^2+(-1)^p\gamma^2\zeta (x)}$.
Again, the ${\mathfrak p}_0$
sign can be easily found: ${\mathfrak p}_{\mu}$ and $\dot x_{\mu}$ are timelike
and ${\mathfrak p}_{\mu}\dot x^{\mu}=-\gamma\sqrt{(-1)^p g}<0$, hence $\sign ({\mathfrak p}_0)=%
-\sign (\dot x_0)$. The result is similar to (\ref{constraint}):
$${\mathfrak p}_0+E_{\mathfrak p}\sign (\dot x^0)=0,$$
the \ "Hamiltonian" is equal to zero and the total Hamiltonian
$H_T=u({\mathfrak p}_0+E_{\mathfrak p}(x, \overrightarrow {\mathfrak p}))+v_i%
{\mathfrak p}_{\mu}x^{\mu}_{,i}$.

Here p+1 momenta are unphysical ones. We assume that
$\det (x_{i,k})\neq 0$ ($x_{i,k}$  is an $p\times p$ matrix) and 
exclude momenta ${\mathfrak p}_1,\ldots, {\mathfrak p}_p$ from
$E_{\mathfrak p}$. Due to (\ref{brone}) we have
${\mathfrak p}_i=([x_{.,.}]^{-1})_{il}({\mathfrak p}_0{x_0}_{,l}-
{\mathfrak p}_{>}x_{> ,l})$. 
Here we denoted all the components
of ${\mathfrak p}^{\mu}$ with $\mu>p$ by the lower index "$_{>}$", and 
$([x_{.,.}]^{-1})_{il}\equiv d_{il}$ is a matrix inverse of $x_{l,i}$.
Then (\ref{brtwo}) turns into quadratic equation
\begin{multline*}
{{\mathfrak p}_0}^2\left(1-d_{il}{x_0}_{,l}d_{ik}%
{x_0}_{,k}\right)
+2{\mathfrak p}_0d_{il}
({\mathfrak p}_{>}x_{> ,l})d_{ik}%
{x_0}_{,k}-\\
-d_{il}({\mathfrak p}_{>}x_{> ,l})d_{ik}({\mathfrak %
p}_{>}x_{> ,k})-(-1)^p\gamma^2\zeta (x)=0.
\end{multline*}
It has two real roots of opposite signs if and only if
$$d_{il}{x_0}_{,l}d_{ik}%
{x_0}_{,k}<1.$$ The sufficient condition is that the norm of
$x_{i,k}$ as a linear operator is greater than the length of
$p$-dimensional vector ${x_0}_{,l}$.
If ${x_0}_{,l}=0$ the simple answer exists:
$$E_{\mathfrak p}=E_{\mathfrak p}(x,{\mathfrak p}_{>})=%
\sqrt{\sum_{i=1}^{p}({\mathfrak p}_i(x, {\mathfrak p}_{>}))^2+
{\mathfrak p}_{>}^2+(-1)^p\gamma^2\zeta (x)}.$$

For the wave function we have path integrals (taking (\ref{condition}) into account)
\begin{multline*}
\psi (x)=\int{\mathit D}^{n+1}{\mathfrak p}{\mathit D}^{n+1}\tilde x
\exp (i[{\mathfrak p}_{\mu}\Delta x^{\mu}-\\
-\omega u({\mathfrak p}_0+E_{\mathfrak p}(\tilde x, {\mathfrak p}_{\perp}))-\omega %
v_i{\mathfrak p}_{\mu}x^{\mu}_{,i}])\psi(\tilde x)=\\
=\int{\mathit D}^{n-p}{\mathfrak p}_{>}{\mathit D}^{n+1}\tilde x
\exp (i[-{\mathfrak p}_{>}\Delta x_{>}-%
\omega uE_{\mathfrak p}(\tilde x, {\mathfrak p}_{>})+
\omega v_i{\mathfrak p}_{>}x_{> ,i}])\times\\
\times\delta (\Delta x^0-\omega u-\omega v_ix^{\prime}_{0,i})%
\prod_{l=1}^r\delta (-\Delta x_l+\omega v_ix_{l,i})\psi (\tilde x).
\end{multline*}

Again $\delta$-functions determine the Lagrange multipliers
\begin{eqnarray*}
\omega v_i=d_{il}\Delta x_l,\\
\omega u=\Delta x^0-\omega v_ix_{0,i}
\end{eqnarray*}
and reduce the number of integrals over $x$:
\begin{multline*}
\psi (x)=\int{\mathit D}^{n-p}{\mathfrak p}_{>}
{\mathit D}^{n-p}{\tilde x}_{>}\exp (i[%
-{\mathfrak p}_{>}\Delta x_{>}-\\
-(\Delta x^0-d_{il}(\Delta x_l)x_{0,i})E_{\mathfrak p}%
(\tilde x, %
{\mathfrak p}_{>})+d_{il}(\Delta x_l)
{\mathfrak p}_{>}x_{> ,i}])\psi(\tilde x).
\end{multline*}

So, we found out the structure of constraints and wrote down the evolution operator
in the same way as in section 2. It isn't surprising remembering results 
of sections 3 and 4. Now we are ready to prove this statement without
fixing any special coordinate system.
We just need to find out the general formula for $\overrightarrow v_{\perp}$.
Notice that by definition ${\mathfrak p}^{\mu}$ lies in a hyperplane of 
$x^{\mu}_{,0}$ and $x^{\mu}_{,i}$.
Then, due to the
constraints (\ref{brone}), ${\mathfrak p}^{\mu}$ is proportional to $x^{\mu}_{\perp ,0}$. We have $v^{\mu}=\frac{\partial x^{\mu}(x^0,\sigma_i)}{\partial x^0}=
\frac{x^{\mu}_{,0}}{x^{0}_{,0}}$ and 
hence ${\mathfrak p}^{\mu}$ is also proportional to $v^{\mu}_{\perp}$:
$v^{\mu}_{\perp}=\alpha {\mathfrak p}^{\mu}$. To find
the coefficient $\alpha$ we take $v^{\mu}_{\perp}{\mathfrak p}_{\mu}=
\frac{x^{\mu}_{\perp ,0}{\mathfrak p}_{\mu}}{x^0_{,0}}=
\frac{-\gamma\sqrt{(-1)^pg}}{x^0_{,0}}$ (the last equality is just the Euler's homogeneous function theorem)
 and ${\mathfrak p}^{\mu}{\mathfrak p}_{\mu}=
(-1)^p\gamma^2\zeta (x)$. But $v^{\mu}_{\perp}{\mathfrak p}_{\mu}=
\alpha{\mathfrak p}^{\mu}{\mathfrak p}_{\mu}$, hence we have
$$v^{\mu}_{\perp}=-\frac{\sqrt{(-1)^pg}}{(-1)^p\gamma\zeta x^0_{,0}}{\mathfrak p}^{\mu}$$
and $1-\overrightarrow v^2_{\perp}=v^{\mu}_{\perp}v_{\mu\perp}=\frac{(-1)^pg{\mathfrak p}^2}
{\gamma^2\zeta^2{x^0_{,0}}^2}=\frac{g}{\zeta {x^0_{,0}}^2}$, so
$$\int dx^0dV\sqrt{1-\overrightarrow v^2_{\perp}}=\int x^0_{,0}\ d\sigma_0 
\sqrt{|\det x^{\mu}_{,i}x_{\mu ,k}|}\ d^p\sigma_i \sqrt{\frac{g}{\zeta {x^0_{,0}}^2}}=
\int d^{p+1}\sigma\sqrt{|g|}.$$
We proved (\ref{setof}) and 
the main statement of sections 3 and 4 for the most general case.

\newpage
{\vspace*{2ex}
{\large \bf \begin{center} References. \end{center}}
\vspace{2ex}}
\begin{enumerate}
\item {\it Krausz F.} Path integral for a scalar propagator: Preprint HUTP
 80-A-003. Cambridge, 1980.
\item {\it Fiziev P.P.} Theor.Math.Phys., {\bf 62}, p. 123 (1985).
\item {\it Prokhorov L.V., Nuramatov A.G.} \quad Vestnik Leningr.
Univ. (Ser. 4) {\bf 3} (18), p. 86 (1991). (in Russian)
\item {\it Prokhorov L.V., Shabanov S.V.} Gamiltonova mekhanika
 kalibrovoch\-nikh sistem.
Izd. SPbGU, 1997. (in Russian)
\item {\it Dirac P.A.M.}  Lectures on quantum mechanics. Dover, 2001.
\item {\it Barbashov B.M., Nesterenko V.V.} Introduction to the relativistic
string theory. World Scientific, 1990.
\item {\it Golovnev A.V., Prokhorov L.V.} \quad
Vestnik SPb Univ. (Ser.4), {\bf 2} (12), p. 86 (2003). (in Russian)
\item {\it Green M.B., Shwartz J.H., Witten E.} Superstring theory, vols. 1 and 2,
Cambridge University Press, 1986.
\item {\it Kaku M.} Introduction to superstrings. Springer-Verlag, 1988.
\item {\it Goto T.} Prog.Theor.Phys, {\bf 46}, p. 1560 (1971).
\item {\it Hara O.} Prog.Theor.Phys, {\bf 46}, p. 1549 (1971).
\end{enumerate}

\end{document}